\documentstyle[11pt]{article}
\everymath{\rm}
\everydisplay{\rm}
\input {amssym.def}
\input {amssym.tex}
\newcommand{\pp}{^{\prime \prime}}

\title{\bf NMR in Copper-Oxide Metals}
\author {C. M. Varma \\
AT\&T Bell Laboratories, Murray Hill, NJ  07974}
\date{}

\begin{document}
\maketitle
\begin{center}
{\bf Abstract}
\end{center}

The anomalous part of the NMR relaxation rate of copper
nuclei in the normal state of copper-oxide metals is
calculated using the orbital magnetic 
parts of the fluctuations derived
in a recent theory to explain the long wavelength
transport anomalies.
Oxygen and Yttrium reside on lattice sites at which the
anomalous contribution is absent at all hole densities.
The frequency, momentum dependence, and the form-factor of the fluctuations is predicted,
which is verifiable by inelastic neutron scattering experiments.
\newpage
Of all the various anomalies in the normal state
of copper-oxide metals,$^1$ the most
extraordinary are the nuclear magnetic
relaxation$^2$ rates $T_1^{-1}$.
These rates probe local fluctuations and are
a very detailed and stringent test of microscopic theories.
The anomalies are essentially identical in the best studied
materials, $YBa_2Cu_3O_{6+x}$ with
$x \approx 0.9$ $(123O_{6.9})$ and
$La_{2- \delta}Sr_{\delta} CuO_4$ with $\delta \approx 0.15$
i.e. near the composition
for the highest $T_c$.
The local magnetic fluctuations in a metal are
usually of the form:
$Im \chi ( \omega ) \sim \frac{\omega N(0)}{E_F}$
leading to $T_1^{-1} \sim T$.
The oxygen $T_1^{-1}$ in the Cu-O metals are
consistent with this behavior,$^3$ as is that
of $Y$ nuclei in (123).$^4$
The Cu-nuclear relaxation rate, on the other hand,
is$^5$ to a first approximation,
$^{Cu}T_1^{-1} \sim a+bT$ with
$a \approx 6T_c$.
The (nearly) constant part in $^{Cu}T_1^{-1}$ suggests
that the local magnetic fluctuations in Cu have a scale-invariant form
at low energies:
$\chi\pp ( \omega ) \sim \omega / T$,$^6$
unlike Fermi-liquids and as if the metals lie near a
$T=0$ critical point.
This is remarkable enough.
But it is even more remarkable that the
local electronic fluctuations on the oxygen do not share
the fluctuations on Cu even though
the electronic wavefunctions on the
two ions are well hybridized.

One idea discussed extensively$^7$ to explain the observations is that
the materials are near an antiferromagnetic instability
of the Cu magnetic moments with a magnetic correlation length
$\xi^{-1} \sim T$.
Since the oxygen ions sit half way between
the coppers, they do not see the antiferromagnetic fluctuations.
Direct measurements$^8$ of $\chi\pp (q, \omega )$ by
inelastic neutron scattering show this idea to be untenable.
In $123O_{6.9}$
the magnetic correlation length is$^{8}$ less than
a lattice constant at low energies and temperature independent.
In $La_{1.85} Sr_{.15} CuO_4$, magnetic correlations
at low energies
are indeed observed.$^9$
But the correlations are at an incommensurate wave-vector so that
the fluctuations do not cancel on the oxygen
ions.
Calculations$^{10}$ of $T_1^{-1}$ using the measured
$\chi\pp (q, \omega )$ give a temperature dependence
which is quite different from the experiments for both
$^{0} T_1^{-1}$ and $^{Cu}T_1^{-1}$.  
Moreover, 
$( ^0 T_1 T K )^{-1} \approx (1.4 \pm 0.2) \sec ^{-1} \deg ^{-1}$
for all measured temperatures in the normal phase for
$123 O_{6 + x}$ for several different
$x$ and for $La_{1.85} Sr_{.15} Cu O_4$.$^{11}$  Here $K$ is the measured
oxygen Knight shift.  This is in agreement with the
value obtained$^{12}$ from the conventional
Korringa Law, in terms of the deduced hyperfine coupling constants.  
The temperature independence of this quantity, as well as its
near invariability from one compound to another is inexplicable by AFM
correlations whose $q$-vector and magnitude change from compound
to compound and with $x$.  Some more robust symmetry is called for so that the
fluctuations seem by Cu nuclei are absent at the oxygen sites.
Clearly, something more subtle is at work.

Recently a model for copper-oxide metals$^{13,14}$ has been
systematically investigated.$^{14}$
A $T = 0$ critical point has been discovered at a unique
composition $x = x_c$, at which the fluctuations lead to
marginal Fermi-liquid self energy:
$Im \Sigma ( \omega ) \sim max ( \omega , T ) \; sgn \omega$,
and long wavelength transport properties in
agreement with experiment.
Here these fluctuations are applied to calculate features of
$T_1^{-1}$.

For reasons discussed elsewhere,$^{13,14}$
the model includes dynamical degrees of
freedom both on Cu and on O ions and includes
long-range interactions
besides a strong short-range repulsion on Cu ions to
exclude the $Cu^{3+}$ state. 
For long-range interaction energies comparable to or larger than the Cu-O bonding -
antibonding splitting, the model has a critical point at $T = 0$, and hole density $x =
x_c$.  Near this point (intracell Cu-O) current fluctuations have a propagator$^{14}$
\begin{equation}
D ( {\bf q}, \omega )= ( 1 / W_s )
( \{i \omega / \gamma + ln(\omega_c / \gamma ) \}^{-1} + A q^2 +
G(x,T)) ^{-1} ,
\end{equation}
Here $W_s$ is the spectral weight of the fluctuations and $\gamma ( \omega , T , G )$,
the damping of this unconserved mode is 
$\approx max (| \omega | , T , G , \tau_0^{-1} )$. 
$\tau_0^{-1}$ is the order of the elastic rate for
single particle scattering from impurities, obtainable, from resistivity measurements.
The "mass" of the collective mode, $G ( x , T ) = g_0 | x - x_c | + T / T_0$,
where $g_0$ is O(1) and $T_0$ is the order of the bandwidth.

We now consider the orbital magnetic contribution of these current modes
to the magnetic fluctuations:
$\chi (q, \omega )$.
First, consider the local fluctuations around
any particular ion with
the Cu-O metal treated in the one-electron approximation.
The local hyperfine Hamiltonian for a
nuclear spin {\bf I} chosen to be at the origin is
\begin{equation}
H_{hf} = {\bf I} \cdot {\bf M} =
{\bf I} \cdot \sum_{kq} {\bf M}_{kq}(0)
c_{k+q/2}^+
c_{k-q/2} ,
\end{equation}
where ${\bf M}_{k,q} (0)$ is the matrix element of the
magnetization at the origin between Bloch waves:$^{15}$
\begin{equation}
|k> = e^{i {\bf k} \cdot {\bf r}}
u_{{\bf k}} ( {\bf r} )
\end{equation}
The contributions to $M$ relevant to us are
\begin{equation}
{\bf M}_{orb} = i \hbar
\frac{{\bf L}}{r^3} =
i \hbar
\frac{{\bf r} \times  {\bf \bigtriangledown}}{r^3} .
\end{equation}

Consider the matrix element
\begin{equation}
\langle {\bf k}+ \frac{{\bf q}}{2}
|{\bf M}_{orb} | {\bf k}-{\bf q}/2 \rangle =
\hbar \langle e^{-{\bf iq} \cdot {\bf r}}
u_{{\bf k}+{\bf q}/2}^* (r)
\left ( \frac { {\bf k} \times {\bf r}}{r^3} + i
\frac{{\bf r}\times {\bf \bigtriangledown}}{r^3} \right )
u_{{\bf k}-{\bf q}/2} (r) \rangle .
\end{equation}

The Bloch-functions
$u_{\bf k}(r)$ are in general very complicated.
We will content ourselves with general properties of (5)
remembering that the relaxation rate sums over all
${\bf k}$ and ${\bf q}$.
To estimate (5) consider $u_{\bf k}$ at some symmetry points where
we know its properties, for example the non-bonding
point ${\bf k} = (0,0)$.
At this point $u_0 ({\bf r})$ can be written as
\begin{equation}
\frac{1}{\sqrt N} \sum_{i,\alpha} a_{i,\alpha}(0) \phi
({\bf r} - {\bf R}_{i,\alpha}),
\end{equation}
where i sums over the cells and $\alpha$ the atoms
in the cell: $\phi ({\bf r} -R_{i,\alpha})$ has the
symmetry of the $d_{x^2-y^2}$ orbitals at
the Cu-site and has the symmetry of the $p_x$ and $p_y$
orbitals, appropriately, at the two oxygen sites per
unit cell.
The phase factors $a_{i,\alpha}(0)$ are $\pm 1$, phased to produce
the non-bonding configuration at ${\bf k} = 0$.
We can verify by ``k.p perturbation'' to determine the phase
factors $a_{i,\alpha}({\bf k})$ for ${\bf k}$ about this
point that the qualitative results remain the same
at arbitrary points in the zone.
So we can use (6) to find the qualitative properties
of (5).

Inserting (6) in (5), we use the rapid fall-off of
$\phi ({\bf r}-{\bf R}_{i\alpha})$ with
$|{\bf r}-R_{i\alpha} |$ to approximate (6) by a
sum of two terms:
(a) $(i,\alpha)$ at the origin (i.e. the position of the
nucleus whose relaxation rate is being evaluated) on both
$u({\bf r})$ factors in (5); (b)
$(i,\alpha )$ at the origin in one $u({\bf r})$ and
$(i,\alpha )=$ nearest neighbors of the atom at the
origin and vice-versa.
In higher order terms the sums over $(i,\alpha )$ may be
replaced by an integral over space.
This is then just the fluctuating electromagnetic field
at the origin due to long wavelength orbital currents.
Such effects have been evaluated$^{16}$ and formed to contribute
$O(10^{-4})$ to the relaxation rates compared to the
on-site relaxation terms (say, from contact hyperfine
interactions) and may be ignored.

The local orbital fluctuation term (a) in (6) is zero at
both Cu and oxygen if as in Cu-O metals non degenerate orbitals
reside at these sites.
Consider now the next term (b).
The second$^{17}$ term in (5)
yields a contribution linear in $q$ given by
\begin{equation}
\langle {\bf k} + \frac{{\bf q}}{2}
| {\bf M}_{orb} | {\bf k} - \frac{{\bf q}}{2} \rangle \simeq
\hbar \sum_n i {\bf q} \cdot
\langle {\bf r} \phi ( {\bf r} - {\bf R}_n)
\frac{{\bf L}}{r^3}
\phi ( {\bf r} ) \rangle ,
\end{equation}
where $n$ sums over the nearest neighbors.
Take a Cu-site at the origin; then $\phi (r)$
has $d_{x^2 -y^2}$ symmetry so that
${\bf L}_z$ $\phi(r) \sim d_{xy}$.
$\phi (r- {\bf R}_n)$ have
$p_x$ or $p_y$ symmetries
so that $\sum_n {\bf q} \cdot {\bf r} \phi ({\bf r}-{\bf R}_n)$
for a general ${\bf q}$
also has a component with $d_{xy}$ symmetry.
Finite matrix element of ${\bf L}_z$ therefore exist.
The construction of a $d_{xy}$ symmetry around a Cu ion by appropriately phasing the
oxygen $p_x$ and $p_y$ orbitals is illustrated in Fig. 1. Semi-classically, orbital relaxation requires a current fluctuation circulating around the nuclei. This is provided by the loop formed by the four oxygens around a given Cu.
These produce magnetization fluctuations in the
z-direction which contribute to the nuclear
relaxation rate $T_{1 \perp}^{-1}$ with field
applied in the Cu-O planes.

The same mechanism also contributes to $^{Cu}T_{1 \parallel}^{-1}$ 
with field applied normal to the
Cu-O plane.  Since $ {\bf L}_{x,y} d_{x^2-y^2} \sim d_{xz} , d_{yz}$ this occurs through
the relative phasing of the oxygen $p_{x,y}$ orbitals in the plane and the $p_z$ orbital
of the apical oxygens with a $ {\bf q}$ in the plane and in general with 
the apical oxygens 
mutually out of phase.

Since $L_z p_x \sim p_y$, etc, orbital relaxation of oxygen nuclei requires a
fluctuation of $p_{x,y}$ form to be created by phasing the neighbors.
For the magnetic fluctuation at the oxygen nuclei, consider
(6) with oxygen sites at the origin.
Oxygen is linearly coordinated by copper ions, and no contribution
of $ O(q)$ exist from the nearest neighbor
terms, because $\sum_n {\bf {q \cdot r}} \phi (r-R_n )$ cannot be
phased to yield a p-symmetry at the oxygen at the origin. 
This is true to any order in q. 
Semiclassically, no simple closed loop current fluctuation can be created around oxygen
unlike the case of copper illustrated in Fig. (1).
Similarly, $Y$ is an s-state ion
sitting between Cu-O planes
and no contributions of the type
we are considering occur for it.

Magnetic fluctuations, which because of the
symmetry of the lattice, contributing only
to relaxation of Cu-nuclear spins therefore do exist.
We now consider their frequency and temperature dependence.
We use the fact that the operators in (4) project on to operators on Cu and
O-orbitals which participate in the current fluctuations of Eq. (3).
The magnetization fluctuations then couple to the current fluctuations through a factor
proportional to ${\bf q}$, as in Eq. (7).
We therefore have an anomalous contribution to the
relaxation at Cu nuclei:
\begin{equation}
T_{1an,i}^{-1} = (T/\omega)
\sum_{\bf q} \langle {\bf M}_{\bf q}
{\bf M}_{-{\bf q}} \rangle_{\omega} = (T/\omega)
\sum_{\bf q}
\alpha_i^2 {\bf q}^2
D ({\bf q} , \omega ) .
\end{equation}
Here $\alpha_i = \alpha_{\perp}$
or $\alpha_{\parallel}$, 
depending on the direction of the magnetic field.
The $\alpha$'s are undetermined numerical factors which depend, besides the couplings in
Eq. (3) on details of the structure and chemistry.
In (8) ${\bf q}$ is two-dimensional;
the magnetization correlations between
the planes is purely from the very small
long-range electro-magnetic effects,
and assumed smaller than the temperature.

Now we note that near $x = x_c$ (ignoring log $T$ corrections)
\begin{equation}
Lim ( {\omega \rightarrow 0})
( T / \omega )
\sum_{{\bf q}} q^2 Im D (q, \omega ) \approx \frac{T}{W_s}
\left [ \frac{1}{\gamma_0 (T)}
+ O ( 1 /  E_F ) \right ].
\end{equation}
Since $\gamma_0 (T) = \tau_0^{-1} + \lambda T$ at G=0,
a scale-invariant contribution
to the local fluctuation on Cu indeed arises
in the pure limit, $\lambda T \gg \tau_0^{-1}$.

We write the ordinary contribution to
nuclear relaxation rate of a nucleus n with
the field applied in the direction i as
\begin{equation}
^nT_{1,i}^{-1} =
lim_{\omega \rightarrow 0}
\sum_{q, \bar{i}}
| ^nF_{\bar{i}} (q_i) |^2
\chi_0\pp (q, \omega ) / \mu_B^2 .
\end{equation}
In (10), $^nF_i (q)$ include the measured hyperfine constants
as well as the lattice form factors, as given, for instance,
in Ref. (18).
${\bar{i}}$ are orthogonal to i.
To this we supplement for Cu the anomalous
contribution coming from (9),
\begin{equation}
^{Cu}T_{1 \; an,i}^{-1} =
\frac{\alpha_{i}^2}{W_s}
\frac{T}{\gamma_0 (T)}  .
\end{equation}

Equation (10) has been evaluated extensively, with
$\chi_0\pp$ calculated in various approximations$^{19,12}$ and with the
form of
$\chi_0\pp (q, \omega )$ deduced from experiments$^{9,10}$ for
$La_{1.86} Sr_{.14} CuO_4$ which show incommensurate
antiferromagnetic correlations.$^{10}$
We defer the role of such correlations,
which in any case are of an ignorable magnitude in $123O_{6.93}$, and
write (10) in the Fermi-liquid form, so that
\begin{equation}
^{Cu}T_{1,i}^{-1} =
\frac{\chi_0}{\mu_B^2 E_F} T
| ^{Cu}A_{\bar{i}} |^2 +
\frac{\alpha_i^2}{W_s}
\frac{T}{\gamma_0 (T)}  ,
\end{equation}
and
\begin{equation}
^0T_{1,i}^{-1} =
\left ( \frac{\chi_0}{\mu_B^2 E_F} \right ) T | ^0A_{\bar{i}} |^2
\end{equation}
where in $^nA_i$, we have absorbed the hyperfine
constants, form factors and other factors after scaling
out a factor $( \chi_0 / \mu_B^2 E_F )$.
Note also that $\alpha_i$
has the same dimensions as $A_i$.

We now quantitatively compare (12) with the experimental results.
We first note that
$lim \; q \rightarrow 0$,
$\omega \rightarrow 0 \; \; Re \; q^2D(q, \omega ) = \alpha^2 /W_s$,
so that for $\alpha^2 \ll 1$, this
provides a negligible contribution to the uniform magnetic
susceptibility and the Knight shift.
So the hyperfine constants deduced from the measured
Knight shift should be used only in the usual
(Fermi-liquid-like) contributions to
$T_1^{-1}$.
At high temperatures, $T \gg \tau_0^{-1}$, (12) behaves
as $a+bT$; the constant part is replaced by the term
proportional to T $\tau_0$ for $T \gg \tau_0^{-1}$.
If three dimensional coupling of the
anomalous magnetization fluctuations were significant,
it would also produce a term proportional to T.

In Fig. (2) we compare the calculated result for
$^{Cu}T_{1, \perp}^{-1}$ from Eq. (12) with field in the plane
with the experimental results in (123$O_{6.93}$).
We deduce
$\frac{\chi_0}{\mu_B^2E_F} |^{Cu} A_{\perp} |^2$ by
scaling up the measured ($T_1T)^{-1}$ of oxygen by the
ratio of the appropriate measured hyperfine coupling constants.
We can estimate $\gamma_0(T)$ from the measured resistivity.
The fit in Fig. (2) is with
$\lambda \approx 1$ and
$\tau_0^{-1} \approx 25K$,
similar to what one deduces from the measured
resistivity of ($1230_{6.93}$).
This leaves the undetermined parameter,
$\alpha_{\perp}^2/ W_s$, to fit the data.
The fit in Fig. (2) is with
$\alpha_{\perp}^2/W_s \approx 3.2 (m sec)^{-1}$,
which with $W_s \approx E_F \approx 10^4K$ implies
$| \alpha_{\perp}/ ^{Cu}A_{\perp}/^2 \approx 6 \times 10^{-2}$.

We also compare the $^{Cu}T_1^{-1}$ measured by NQR in a single
crystal of $La_{1.86}Sr_{.14}CuO_4$.
The only significant difference in parameters needed to fit
the data is a larger value
$\tau_0^{-1} \approx 10^2K$.
This is consistent with the fact that the extrapolated
residual resistivity of optimum (124) crystals is generally
higher than that of optimum (123) crystals.
The fit in Fig.\ (2) uses
$\alpha_{\perp}^2 / W_s \approx 2.8 (m sec)^{-1}$.

To see if $\alpha_{\perp}^2$ is reasonable, we
compare it to a hypothetical situation where
$d_{x^2 -y^2}$ and $d_{xy}$ orbitals on Cu
are degenerate, so that a {\it local} orbital as well
as a dipolar fluctuation contribution to
$T_1^{-1}$ exists with field in the plane.
The ratio $|A_{orb} / A_{\perp}|^2$
in this case (where both relaxation rates are $\sim T$)
can be easily calculated$^{20}$ and is $\approx 20$.
We have two reduction factors
in $\alpha^2$ compared to $A_{orb}^2$,
that due to
$\langle \phi_i^2 \rangle \sim x$, and
more importantly, because the distance to
the neighboring oxygen p-orbitals is involved.
A reduction factor of $O(10^{-2}$) is therefore not unreasonable.

In general $( \alpha_{\perp} / \alpha_{\parallel} )^2 > 1$ is
expected, the details depending on structure.  Experimentally this ratio is about 2.6 in
$La_{1.85} Sr_{.15} Cu O_4$ and about 4 in $Y Ba_2 Cu_3 O_{6.9}$.
This is in the right direction from the consideration here since the former is a single
layer and the latter a double layer compound.  
Measurements of anisotropy in other compounds appear not to have been made.
We would predict that this ratio rises in compounds with larger number of layers.
Any numerical estimates do not appear feasible.

At compositions away from the ideal, i.e.
$g_0 | x - x_c | >> T / T_0$,
the fluctuation modes of eq. (1) have a gap.
This has the same effect in $T_1^{-1}$ as an enhanced
$\tau_0^{-1}$.
This is qualitatively consistent with the experimental
results that there is a smoother variation of
$T_1^{-1}$ from low temperatures to high away from the
ideal composition.
It is worth noting also that a finite G also leads to a cross-over
at temperatures of O(G) from a lower effective
mass to a higher effective mass.
This should be observable in thermodynamic properties: both specific
heat and magnetic susceptibility as indeed it is.$^{21}$

As has been discussed before,$^{12,19}$ the
$^OT_1^{-1}$ in $(123O_7)$ is consistent with (13)
and given quantitatively in terms of the measured
$\chi_0$ and hyperfine constants. This is true in the present theory at all x.
For $^0T_1^{-1}$ in $La_{1.85} Sr_{.15} Cu O_4$ to be consistent with the neutron
scattering results, the {\em absolute magnitude} of the q-dependent part of
$\chi\pp (q , \omega )$ at low $\omega$ should be about a factor of 3 less than used in
Ref. (10) or the deduced (hyperfine constants)$^2$ smaller by a similar factor.

In contrast to $T_1^{-1}$ which is related to fluctuations
integrated over ${\bf q}$, $T_2^{-1}$ of Cu, as derived by
Pennington and Slichter$^2$ is dependent primarily to fluctuation
at $q \approx \pi /R_{Cu-Cu}$.
So antiferromagnetic correlations, however weak are
picked up in $T_2^{-1}$.
Berthier et al.$^{11}$ find using the
neutron scattering results
as a function of temperature in 123$O_{6.7}$ and in 123$O_{6.93}$
that the measured correlation length of
only about two lattice constants
in the former and one or less in the latter accounts for
the measured $T_2^{-1}(T)$.
Similar conclusions have been arrived at by Walstedt for
$La_{1.86}Sr_{.14}CuO_4$.

>From Eqs. (1), (4) and (7) one can deduce that the orbital magnetic fluctuations
contribute
\begin{equation}
Im \chi_{orb} ( q , \omega ) \backsimeq
\mu_B^2
\left ( \frac{a_0}{a} \right )^6
( a^2 q^2 ) Im \: D (q , \omega )
\end{equation}
to the total magnetic fluctuations, $Im \chi ( q, \omega )$, 
with a form factor such that they are seen only at the Cu (or symmetry equivalent)
sites and absent at the oxygen (or symmetry equivalent) sites.
Here $a \approx 1.9 \AA$ is the nearest neighbor Cu-O distance, and
$a_0$ is of the order of the atomic radius. 
To this should be added the usual spin-fluctuation contributions.
Should there be significant (but not singular) AFM correlations, 
for example due to nesting, the
Fermi-liquid contributions to Eqs. (12) and (13) can easily be modified to include their
effect.

A contribution, smooth in both $q$ and
$\omega$ extending up to high energies, as in Eq. (14) is indeed observed.$^{22}$  To test
Eq. (14) in detail requires measurements up to high frequencies at various temperatures.
The only measurement reported$^{22}$ up to high frequencies has been done only at $T =
17K$ in $La_{1.85} Sr_{.15} Cu O_4$.  Fig. (3) of Ref. (22) gives the frequency
dependence of the integrated absorption as very slowly decreasing to $\sim 0.2$ eV below
a weak peak at $\sim 20$ meV.  This is quite consistent with Eq. (14).  Measurements at
other temperatures in that compound and up to high frequencies and various temperatures
in $123 O_{6.9}$ are urged.
Especially crucial are measurements of the fluctuation spectrum in several Brillouin
zones to deduce separately the fluctuations centered on Cu alone and on O alone 
The former should see (14) while the latter should be the usual
Fermi-liquid form, i.e. $\approx N (0) \omega / q V_F$
for $\omega \lesssim q V_F$ and $\approx 0$ beyond.  These are
hard experiments, but they should resolve the principal problem
in the field.

I am pleased to acknowledge very helpful discussions with
A. Sengupta, G. Kotliar, P. B. Littlewood and Q. Si about
this work, and extensive discussions on NMR with R. E. Walstedt,
H. Alloul, C. Berthier and C. Hammel and on neutron scattering
with G. Aeppli.
\newpage
\begin{center}
{\bf References}
\end{center}
\begin{enumerate}
\item For a review, see B. Batlogg in
{\it High Temperature Superconductivity}, ed. by
K. Bedell et al., Addison Wesley, New York (1989).
\item For review, see C. H. Pennington and C. P. Slichter
in {\it Physical Properties of High Temperature Superconductors II},
D. M. Ginsburg, editor, World Scientific, Singapore (1989).
\item P. C. Hammel et al., Phys. Rev. Lett. {\it 63},
1992 (1989)
\item H. Alloul, T. Ohno and P. Mendels, Phys. Rev. Lett. {\it 63},
1700 (1989)
\item R. E. Walstedt et al., Phy. Rev. B {\it 38}, 9299 (1988).
\item C. M. Varma et al., Phys. Rev. Lett. {\it 63}, 1996 (1989).
\item B. S. Shastry, Phys. Rev. Lett. {\it 63}, 1288 (1989).
\item J. Rossat-Mignod et al., Physica {\it 192 B}, 109 (1993);
J. M. Tranquada et al. Phys. Rev. {\it B46}, 5561 (1992);
H. A. Mook et al., Phys. Rev. Lett. {\it 70}, 3490 (1993);
L. P. Regnault et al., Physica C {\it 235}, 59 (1994);
B. Keimer et al., (preprint).
\item T. E. Mason et al., Phys. Rev. Lett. {\it 71}, 919 (1993).
\item R. E. Walstedt, B. S. Shastry and S.-W. Cheong,
Phys. Rev. Lett. {\it 72}, 3610 (1994).
It is also worth noting that the result reported (T. Imai, Phys. Rev. Lett.
{\it 70}, 1002 (1993)) that at high temperatures
$La_{1.85} Sr_{.15} Cu O_4$ has the same $^{Cu} T_1^{-1}$ as
$La_2 Sr Cu O_4$ has been modified by more recent measurements
(M. Matsumra et al., J. Phys. Soc. Japan {\it 65}, 699 (1996)).
\item C. Berthier et al., Physica C {\it 185-9}, 1141 (1991);
Physica Scripta {\it T49}, 131 (1993)
and private communication; R. Walstedt (unpublished).
\item P. B. Littlewood et al., Phys. Rev. B {\it 48}, 487 (1993).
\item C. M. Varma, S. Schmitt-Rink and E. Abrahams, Solid State
Comm. {\it 62}, 681 (1987).
\item C. M. Varma, Phys. Rev. Letters. {\it 75}, 898 (1995) and Preprint (1996),
submitted to Physical Review and available on http://www.lanl.gov as cond-mat/9607105.  
In this
preprint, some important modifications of the results in the PRL are given.  In
particular, a damping mechanism of the collective mode due to particle-hole scattering
near the Fermi-surface leads to the form of $\gamma ( \omega , T)$ in Eq. (1) which is
more important than Landau damping for all momenta.
\item In the tight-binding approximation the phase
factors for electronic wave-functions is given only
at the lattice points.
The results derived below depend on varying the phase
of the wave-functions between lattice points.
Therefore Bloch waves must be used for the electronic
wave-functions.
\item P. A. Lee and N. Nagaosa, Phys. Rev. {\it 43}, 1223 (1991).
\item For general k's there is also a contribution from the
first term in (5) with properties similar
to those evaluated here.
Since we are unable to evaluate any of the matrix elements
quantitatively and the aim is only to show the feasibility
of the idea and fit to experiment with reasonable parameters,
we do not discuss these.
\item (a) F. Mila and T. M. Rice, Phys. Rev. B {\it 40}, 11382 (1989);
(b) A. J. Millis and H. Monien, Phys. Rev. B {\it 45}, 3059 (1992).
\item Q. Si, Y. Zha, K. Levin and J. P. Lu, Phys. Rev. {\it 47},
9055 (1993) and references therein.
\item Y. Obata, J. Phys. Soc. Japan {\it 18}, 1020 (1963).
\item J. W. Loram et al., Physica C {\it235-240}, 134 (1994).
\item S. M. Hayden et al., Phys. Rev. Lett., {\it 76}, 1344 (1996).
\end{enumerate}
\newpage
\begin{center}
{\bf Figure Captions}
\end{center}

{\it Figure 1:}
Illustrates the physical content of Eq. (10).  A Cu $d_{x^2 - y^2}$ orbital is surrounded
by the $p_{x,y}$ orbitals of oxygen atoms with the phase as shown.  A deviation of the
phase $\sim {\bf q}$ in the direction shown produces a fluctuation with $d_{xy}$
symmetry (shown shaded) about the Cu-site leading to a orbital magnetic moment
fluctuation $\sim {\bf q}$.

{\it Figure 2:}
Experimental results from Ref. (5) for Cu nuclear relaxation rate with field the planes
in $YBa_2 Cu_3 O_{6.9}$, and nuclear quadrupole relaxation rate in
$La_{1.85} Sr_{.15} Cu O_4$ from Ref. (10) compared with the calculations in this paper.
The two parameters required are discussed in the text.
\enddocument